\documentclass{amsart}

\usepackage{amsmath}
\usepackage{mathtools}
\usepackage{amssymb}
\usepackage{enumerate}
\usepackage{slashed}
\usepackage{graphicx}
\usepackage{newlfont}
\usepackage{amsrefs}
\usepackage{comment}
\usepackage[normalem]{ulem}
\usepackage{mathtools}
\usepackage{accents}
\usepackage{leftidx}

\numberwithin{equation}{section}

\theoremstyle{plain}
\newtheorem{theorem}{Theorem}%[section]

\theoremstyle{definition}

\theoremstyle{remark}

\renewcommand{\div}{\operatorname{div}}

\newcommand{\R}{\mathbb{R}}

\newcommand{\D}{\mathcal{D}}

\newcommand{\Z}{\mathbb{Z}}
\newcommand{\M}{\mathcal M}

\newcommand{\dist}{\operatorname{dist}}

\begin{document}

\title[ALE/AKK Black Holes in 5 Dimensions]{Asymptotically Locally Euclidean/Kaluza-Klein
Stationary Vacuum Black Holes in 5 Dimensions}

\author[Khuri]{Marcus Khuri}
\address{Department of Mathematics\\
Stony Brook University\\
Stony Brook, NY 11794, USA}
\email{khuri@math.sunysb.edu}

\author[Weinstein]{Gilbert Weinstein}
\address{Physics Department and Department of Mathematics\\
Ariel University\\
Ariel, 40700, Israel}
\email{gilbertw@ariel.ac.il}

\author[Yamada]{Sumio Yamada}
\address{Department of Mathematics\\
Gakushuin University\\
Tokyo 171-8588, Japan}
\email{yamada@math.gakushuin.ac.jp}

\thanks{M. Khuri acknowledges the support of
NSF Grant DMS-1708798. S. Yamada acknowledges the support of JSPS grants KAKENHI 24340009 and 17H01091.}

\begin{abstract}
We produce new examples, both explicit and analytical, of bi-axisymmetric stationary vacuum black holes in 5 dimensions. A novel feature of these solutions is that they are asymptotically locally Euclidean in which spatial cross-sections at infinity have lens space $L(p,q)$ topology, or asymptotically Kaluza-Klein so that spatial cross-sections at infinity are topologically $S^1\times S^2$. These are nondegenerate black holes of cohomogeneity 2, with any number of horizon components, where the horizon cross-section topology is any one of the three admissible types: $S^3$, $S^1\times S^2$, or $L(p,q)$.
Uniqueness of these solutions is also established.
Our method is to solve the relevant harmonic map problem with prescribed singularities,
having target symmetric space $SL(3,\mathbb{R})/SO(3)$. In addition, we analyze the possibility of conical singularities and find a large family for which geometric regularity is guaranteed.
\end{abstract}
\maketitle

\section{Introduction}
\label{introduction}

The study of higher dimensional ($D>4$) black holes has received substantial interest in recent years, primarily motivated by considerations in string theory. Some intriguing features of these objects, which separate them from their 4-dimensional counterparts, include nontrivial horizon topologies and failure of the classical no hair theorem \cite{EmparanReall}. Another curious attribute that has remained relatively unexplored is the possibility of nonstandard asymptotics, that is, the presence of ends which are not asymptotically flat (approaching Minkowski space $\mathbb{R}^{D-1,1}$) or asymptotically Kaluza-Klein (approaching $\mathbb{R}^{d,1}\times T^{D-d-1}$ where $T^{D-d-1}$ is a torus). Trivial examples of such vacuum spacetimes may be constructed from the Schwarzschild-Tangherlini solution by replacing the round spheres $S^{D-2}$, which foliate the constant time slices, by quotients $S^{D-2}/\mathcal{G}$ where $\mathcal{G}$ is a discrete subgroup of the group of isometries $O(D-1)$. In this case the spacetime is asymptotic to $\mathbb{R}^{D-1,1}/\mathcal{G}$, and is neither asymptotically flat nor asymptotically Kaluza-Klein; its horizon cross-section has topology $S^{D-2}/\mathcal{G}$. Spacetimes with these asymptotics may be referred to as asymptotically locally Euclidean (ALE) \cite{ChenTeo} or asymptotically locally flat (ALF) \cite{HollandsIshibashi}, although the later terminology is not consistent with definitions commonly used in the mathematics literature \cite{ChenChen}.

In \cite{LuMeiPope} L\"{u}, Mei, and Pope constructed explicit solutions of the stationary vacuum Einstein equations in $D=5$ with a bi-axisymmetry. Thus, in contrast to the cohomogeneity 1 quotients of the Schwarzschild-Tangherlini spacetime above, these black holes are cohomogeneity 2. Moreover, in the static limit both the horizon topology and that of the spatial sections at infinity are the lens space $L(p,q)=S^{3}/\mathbb{Z}_p$, where $p$ and $q$ are relatively prime positive integers. These solutions are then ALE with nontrivial horizon topology. The general stationary metrics are characterized by four parameters, namely, two independent angular momenta and one mass as in the Myers-Perry solution, and one extra parameter that plays a role similar to a NUT charge. These solutions were found following a method analogous to the procedure employed in the construction of general type D metrics in four dimensions  \cite{Plebanski,PlebanskiDemianski}. See also \cite{ChenTeo} for other examples of ALE spacetimes.

The purpose of the current paper is to introduce a new technique for constructing ALE as well as asymptotically Kaluza-Klein (AKK) stationary vacuum black holes in 5D. The methodology is versatile in that it allows for any combination of admissible horizon topology and spatial section at infinity to coexist within the same spacetime. For example, a lens $L(p,q)$ horizon may be combined with a ring $S^1\times S^2$ spatial section at infinity to yield an AKK lens black hole, or a ring horizon may be combined with a lens spatial section at infinity to produce a ALE black ring. Furthermore, we are also able to assemble multiple component horizons of differing topologies, again with various spatial sections at infinity.
Our approach is based on solving the harmonic map problem with prescribed singularities which naturally arises from the dimensional reduction procedure for stationary bi-axisymmetric vacuum black holes. It is an extension of the work in \cite{KhuriWeinsteinYamada}, which treated the traditional asymptotically flat case. More precisely, we will prove existence and uniqueness of an axially symmetric harmonic map $\Phi\colon \R^3\setminus\Gamma\to SL(3,\R)/SO(3)$ with prescribed singularities on a subset $\Gamma$ of the $z$-axis. The type of prescribed singularities determines not only the black hole topology but the nature of the asymptotics at spatial infinity as well.

%In~\cite{khuriweinsteinyamada2017}, we used the reduction of the Einstein vacuum %equations in 5-d to construct asymptotically flat stationary bi-axially symmetric %black hole solutions. In this continuation paper, we consider the same problem in %the asymptotically locally Euclidean (ALE) setting. The asymptotic infinity is then %a $\R\times M$ where $M$ is 3-manifold of positive Yamabe type. As %in~\cite{khuriweinsteinyamada2017} the problem reduces to an axially symmetric %harmonic map $\Phi\colon \R^3\setminus\Gamma\to SL(3,\R)/SO(3)$ with prescribed %singularities on $\Gamma$ a subset of the $z$-axis. The target $SL(3,\R)/SO(3)$ of %the harmonic map is a symmetric space of non-compact type and rank 2. In fact, the %proof follows the same lines, with the only difference being the prescription of %the behavior at infinity.

\section{Background and Statement of Main Results}
\label{background}

Consider a stationary vacuum bi-axisymmetric 5-dimensional spacetime $\mathcal{M}^5$, and let $\partial_{t}$, $\partial_{\phi^i}$, $i=1,2$ denote the generators of the symmetry group $\mathbb{R}\times U(1)^2$. Under mild hypotheses, the orbit space of the domain of outer communication $\mathcal{M}^5/[\mathbb{R}\times U(1)^2]$ is known \cite{HollandsYazadjiev1} to be homeomorphic to the right half plane $\{(\rho,z)\mid \rho>0\}$. This may be enhanced to cylindrical coordinates $(\rho,z,\phi)$ on $\mathbb{R}^3\setminus \{z-\text{axis}\}$, which plays the role of domain for the relevant axisymmetric harmonic map. The boundary $\rho=0$ of the right half plane contains information concerning the topology of the horizon as well as the asymptotic structure at infinity. In this regard the $z$-axis is broken into intervals called \textit{rods}
\begin{equation}
\Gamma_1=[z_1,\infty),\text{ }\Gamma_2=[z_2,z_1],\text{ }\ldots,\text{ }
\Gamma_{L}=[z_{L},z_{L-1}],\text{ }\Gamma_{L+1}=(-\infty,z_{L}].
\end{equation}
On each interval a linear combination $m_l \partial_{\phi^1}+n_l \partial_{\phi^2}$ vanishes, where $m_l$ and $n_l$ are integers which are relatively prime whenever both are not zero. The tuple $(m_l,n_l)$ is referred to as the \textit{rod structure} of the rod $\Gamma_l$. A \textit{horizon rod} is an interval on which no closed-orbit Killing field degenerates, that is $m_l=n_l=0$, and the remaining intervals are \textit{axis rods}. End points of horizon rods are called \textit{poles}, whereas the remaining interval end points are \textit{corners}.

The topology of a horizon component associated to a horizon rod $\Gamma_l$ may be identified from the rod structure as follows. Connect the adjacent rods via a semicircle in the right half plane starting at $\Gamma_{l-1}$ and ending on $\Gamma_{l+1}$, and enclosing $\Gamma_l$. By turning on the $U(1)^2$ symmetry, each point on the interior of this semicircle represents a 2-torus, and at each end point a 1-cycle of the torus degenerates. Thus, we obtain a 3-manifold with a singular foliation by tori, and the topology is determined by which 1-cycles collapse at the end points. For example if $\Gamma_{l-1}$, $\Gamma_{l+1}$ have rod structures $(1,0)$, $(0,1)$ we obtain a sphere $S^3$, whereas $(1,0)$, $(1,0)$ yields a ring $S^1\times S^2$, and $(1,0)$, $(q,p)$ produces a lens $L(p,q)$. Similarly, by foliating infinity in the orbit space by such semicircles connecting the two semi-infinite rods $\Gamma_1$ and $\Gamma_{L+1}$, the end within a constant time slice has topology $\mathbb{R}\times S^3$ (asymptotically flat), $\mathbb{R}\times S^1\times S^2$ (AKK), and $\mathbb{R}\times L(p,q)$ (ALE) respectively.

Let $p_l$ be a point on the $z$-axis where a corner is present, and let
$(m_l,n_l)$ and $(m_{l+1},n_{l+1})$ be the rod structures for the surrounding axis rods.
In order to prevent orbifold singularities, the \emph{admissibility condition}
\begin{equation} \label{0192}
\det\begin{pmatrix} m_l & n_l \\ m_{l+1} & n_{l+1} \end{pmatrix} = \pm 1
\end{equation}
is imposed. A further hypothesis, referred to as the \textit{compatibility condition} will be needed for technical reasons arising from the harmonic map existence proof. To state this condition, let $p_{l-1}$ and $p_{l}$ be two consecutive corners, flanked by axis rods $\Gamma_{l-1}$, $\Gamma_{l}$, and $\Gamma_{l+1}$. It may be assumed without loss of generality that the determinants \eqref{0192} associated with the corners $p_{l-1}$ and $p_{l}$ are both $+1$. We then require
\begin{equation}\label{1029}
m_{l-1}m_{l+1}\leq 0.
\end{equation}

As is well-known \cite{HollandsIshibashi,IdaIshibashiShiromizu} the stationary bi-axisymmetric vacuum Einstein equations reduce to solving the harmonic map equations
\begin{align}\label{harmonicmap}
\begin{split}
\tau^{f_{lj}}:=&\Delta f_{lj}-f^{km}\nabla^{\mu}f_{lm}\nabla_{\mu}f_{kj}
+f^{-1}\nabla^{\mu}\omega_{l}\nabla_{\mu}\omega_{j}=0,\\
\tau^{\omega_{j}}:=&\Delta\omega_{j}-f^{kl}\nabla^{\mu}f_{jl}\nabla_{\mu}\omega_{k}
-f^{lm}\nabla^{\mu}f_{lm}\nabla_{\mu}\omega_{j}=0,
\end{split}
\end{align}
where the vector $\tau$ denotes the tension field, $F=(f_{ij})$ is a $2\times 2$ symmetric positive definite matrix determining the rod structure, $f=\det F$, and $\omega=(\omega_1,\omega_2)^{t}$ are twist potentials. The spacetime metric on $\mathcal{M}^5$ associated with these quantities is given in Weyl-Papapetrou coordinates by
\begin{equation}\label{metric}
g=f^{-1}e^{2\sigma}(d\rho^2+dz^2)-f^{-1}\rho^2 dt^2
+f_{ij}(d\phi^{i}+v^{i}dt)(d\phi^{j}+v^{j}dt),
\end{equation}
where the $v^i$ are obtained from the twist potentials by quadrature. From this we see that the rod structure may be interpreted as a vector $(m_l,n_l)^t$ lying in the (1-dimensional) kernel of the matrix $F$ at an axis rod $\Gamma_l$.

Let $\Gamma$ be the union of all axis rods, then the relevant harmonic map $\Phi:\mathbb{R}^3\setminus\Gamma\rightarrow SL(3,\mathbb{R})/SO(3)$ may be constructed from $(F,\omega)$ and represented as a $3\times 3$ symmetric positive definite unimodular matrix \cite{Maison}. Boundary conditions for the potentials $\omega$ are given by constants $\mathbf{c}_l\in\R^2$ on each axis rod $\Gamma_l$, such that the values of the constants are the same on consecutive axis rods. Thus, these constants only change value across a horizon rod, and the difference is proportional to the angular momenta of the associated horizon component. We define a \textit{rod data set} $\D$ to be the rods $\{\Gamma_l\}$ with rod structures $\{(m_l,n_l)\}$, and the potential constants $\{\mathbf{c}_l\}$.

%An asymptotically flat stationary vacuum spacetime will be referred to as %\textit{well-behaved} if the
%orbits of the stationary Killing field are complete, the
%domain of outer communication (DOC) is globally hyperbolic, and the DOC
%contains an acausal spacelike connected hypersurface which is asymptotic to the canonical %slice in the asymptotic end and whose boundary
%is a compact cross section of the horizon. These assumptions are consistent with those of %\cite{ChruscielCosta}, and are used for the reduction of the stationary vacuum equations. %The main result may now be stated as follows.

\begin{theorem} \label{thm1}\par
Given a rod data set $\mathcal{D}$ respecting the admissibility and compatibility conditions, there exists a unique harmonic map $\Phi=(F,\omega)\colon\R^3\setminus\Gamma\to SL(3,\R)/SO(3)$ which realizes the prescribed potential constants and rod structures of $\mathcal{D}$. From this a stationary vacuum bi-axisymmetric black hole spacetime may be constructed with prescribed angular momenta, in which the topology of each horizon component and spatial cross-section at infinity is prescribed to be either $S^3$, $S^1\times S^2$, or $L(p,q)$.
\end{theorem}

This is analogous to the main result of \cite{KhuriWeinsteinYamada} which treated the
asymptotically flat case. Here, in contrast, asymptotically Kaluza-Klein and asymptotically locally Euclidean black holes are produced where the spatial cross-sections at infinity are $S^1\times S^2$ and $L(p,q)$, respectively.
These solutions do not possess closed timelike curves due to the nature of their construction using the Weyl-Papapetrou form of the metric.
It should also be noted that Theorem \ref{thm1} yields essentially all possible black holes of this type.
Two issues which are not immediately answered by the theorem are the questions of analytic regularity of the metric coefficients across the axes, and the possibility of conical singularities. Nevertheless, in Sections \ref{examples} and \ref{conical} we are able to demonstrate that Theorem \ref{thm1} produces geometrically regular (having no conical singularity) ALE black lenses and AKK black rings. These are the first geometrically regular solutions to be constructed using a PDE approach.

\begin{theorem} \label{thm2}\par
The solutions produced in Theorem \ref{thm1} have no conical singularity at spatial infinity. In particular, the two semi-infinite rods $\Gamma_1$ and $\Gamma_{L+1}$ are always void of conical singularities.
\end{theorem}

It is worthwhile to point out that this theorem applies to the asymptotically flat case
treated in \cite{KhuriWeinsteinYamada}, as this falls within the ALE setting.

\section{Examples} \label{examples}

In this section we construct two classes of explicit solutions. The first is a rotating
lens black hole with an asymptotically locally Euclidean end, and the second is a static (non-rotating) ring black hole with an asymptotically Kaluza-Klein end. Both classes of spacetime will be used to construct model maps which play an important role in the proof of Theorem \ref{thm1}.
The static example will be promoted to a fully rotating ring solution by application of Theorem \ref{thm1}, and will be shown to have no conical singularities in Section \ref{conical}. The black lens examples will be geometrically regular by virtue of their construction.

\subsection{Quotients of Myers-Perry} \label{MP}

Let $\M^5$ be a Myers-Perry black hole \cite{MyersPerry}. This is a 3-parameter family of spherical black holes, parameterized by mass and two angular momenta. They are asymptotically flat stationary vacuum solutions with a bi-axisymmetry. Thus, the subgroup
$\mathbb{Z}_p\subset U(1)^2$ acts on $\mathcal{M}^5$ by isometries of the form $\phi_1\rightarrow \phi_1 + 2\pi/p$, $\phi_2\rightarrow \phi_2+2\pi q/p$, where $p$ and $q$ are relatively prime positive integers. Since this action is properly discontinuous the quotient $\M^5/\Z_p$ is a smooth manifold, and with the quotient metric it is a solution of the vacuum Einstein equations. Cross-sections of the horizon now have the topology of the lens space $L(p,q)$, and the asymptotic region has the topology  $\R\times\R\times L(p,q)$, making these solutions ALE.

The quotient spacetime has three rods
\begin{equation}\label{rods}
\Gamma_1=[a,\infty),\quad \Gamma_2=[-a,a],\quad \Gamma_3=(-\infty,-a],
\end{equation}
and the corresponding rod structures are $(1,0)$, $(0,0)$, $(q,p)$. Note that the northern most and southern most rods are axes, while the middle rod denotes a nondegenerate horizon for $a>0$. Dimensional reduction yields a singular harmonic map
$\Phi\colon\R^3\setminus\Gamma \to SL(3,\R)/SO(3)$. These solutions are all regular.

%Consequently $\tau(\Phi)=0$. We note that there is such a solution for any value the mass %and of the two rod potential constants, i.e.\ for any value of the two angular momenta. %In particular such a quotient of the static Schwarzschild-Tangherlini solution exists. %These solutions are all regular.

A special case of these examples are the quotients of the Schwarzschild-Tangherlini black holes mentioned in the introduction. Moreover in the limiting case where the potential constants vanish and $a\rightarrow 0$, we obtain the quotients of Minkowski space by the same $\Z_p$ subgroups. Such solutions will have an orbifold singularity at the origin, but will be regular elsewhere. It is not difficult to find the corresponding harmonic map $\Phi_{\text{lens}}=(F_{\text{lens}},\mathbf{0})$, namely
\begin{equation}\label{ALEF}
	F_{\text{lens}} = h \begin{pmatrix} r\sin^2(\theta/2) & 0 \\ 0 & r\cos^2(\theta/2) \end{pmatrix} h^t, \quad\quad\quad
	h=\begin{pmatrix} 1 & 0 \\ -q/p & 1/p \end{pmatrix},
\end{equation}
where $(r,\theta)$ are polar coordinates in the half plane in which $\theta=0$ corresponds to the positive $z$-axis. These expressions will be used in the construction of model maps near infinity in the next section.

\subsection{Ring-Like Infinity} \label{ring}

Consider the same set of three rods \eqref{rods} and define
\begin{equation}\label{ringF}
		F_{\text{ring}}=\begin{pmatrix} e^{u} & 0 \\ 0 & 1 \end{pmatrix},\qquad
u=u_{a} + v_{-a},
\end{equation}
where
\begin{equation}
	u_a =\log(r_a-(z-a))=\log\bigl(2r_a\sin^2(\theta_a/2)\bigr), \quad\quad
v_a = \log(r_a+(z-a))=\log\bigl(2r_a\cos^2(\theta_a/2)\bigr).
\end{equation}
Here $(r_a,\theta_a)$ denotes polar coordinates in the half plane centered at the point $z=a$ on the $z$-axis. Since the functions $u_a$ and $v_a$ are harmonic, and hence so is $u$, it follows that $\Phi_{\text{ring}}=(F_{\text{ring}},\mathbf{0})$ is a harmonic map.  The rod structure from north to south is $(1,0)$, $(0,0)$, $(1,0)$. The horizon topology is then that of the ring $S^1\times S^2$, and the asymptotic region has the topology $\R\times\R\times S^1\times S^2$. Therefore this gives rise to a static AKK black ring spacetime. These will also be used in the construction of model maps in the next section.

In order to construct the spacetime metric \eqref{metric}, it remains to find $\sigma$.
After a series of standard calculation \cite{KhuriWeinsteinYamada} we get
\begin{equation}\label{equation11}
\partial_\rho \sigma = - \rho T_{zz},  \qquad \partial_z \sigma = \rho T_{\rho z},
\end{equation}
where
\begin{equation}
T_{ij} dx^i \otimes dx^j= (d \rho \,\, dz)  \begin{pmatrix} \frac14 (\partial_{\rho} u)^2 - \frac14 (\partial_{z} u)^2  & \frac12
\partial_{\rho} u \partial_{z} u  \\
\frac12 \partial_{\rho} u \partial_{z} u & - \frac14 (\partial_{\rho} u)^2 + \frac14 (\partial_{z} u)^2 \end{pmatrix}
\begin{pmatrix} d \rho \\ dz \end{pmatrix}  \qquad \quad x^1=\rho, \quad x^2=z.
\end{equation}
The harmonic map equations guarantee that the right-hand sides of \eqref{equation11} form a closed 1-form, guaranteeing the existence of $\sigma$. Using the formula
\begin{equation}
u = \log\left( [\sqrt{\rho^2 + (z-a)^2} -(z -a)][\sqrt{\rho^2 + (z+a)^2} +(z +a)]\right)
\end{equation}
an expression for $\sigma$ may then be found by integrating the differential $d\sigma$.
In the limiting case when $a\rightarrow 0$, we have $u = 2 \log \rho$ and $\sigma= \log \rho$ so that the spacetime is $\mathbb{R}^{3,1}\times S^1$ with flat metric
\begin{equation}
g =  - dt^2 + \rho^2 (d \phi^1)^2 +  (d \phi^2)^2 + d\rho^2 + dz^2.
\end{equation}

\section{The Model Maps}
\label{model}

In this section we construct the model map $\Phi_0:\mathbb{R}^3\setminus \Gamma\rightarrow SL(3,\mathbb{R})/SO(3)$ used to prescribed the singular behavior of the desired harmonic map $\Phi$ near the axis $\Gamma$, as well as the asymptotics at infinity. The requirement on this map is that it has uniformly bounded tension $|\tau(\Phi_0)|<C$, and decays appropriately $|\tau(\Phi_0)|=O(r^{-\alpha})$, $\alpha>2$. The harmonic maps produced in Theorem \ref{thm1} will be \textit{asymptotic} to the model map in the sense that the distance $d(\Phi,\Phi_0)$ in the target $SL(3,\mathbb{R})/SO(3)$ will remain bounded near $\Gamma$, and the distance will asymptote to zero at infinity. The model map may be thought of as an approximate solution to the singular harmonic map problem on which the exact solution will be built.

Before proceeding with the construction, we first collect a few formulas from \cite{KhuriWeinsteinYamada} which are needed to aid computations. Using the same parameterization $(F,\omega)$ of the target manifold as described in the previous section, the symmetric space metric takes the form
\begin{equation}
  \frac14 \frac{df^2}{f^2} +\frac14 f^{ij}f^{kl}df_{ik}df_{jl}  + \frac12 \frac{f^{ij} d\omega_i d\omega_j}{f}
  	= \frac14 [\mathrm{Tr}(F^{-1}dF)]^2  +\frac14 \mathrm{Tr}(F^{-1}dF\, F^{-1} dF)  + \frac12 \frac{d\omega^t \, F^{-1}\, d\omega}{f},
\end{equation}
where $f=\det F$. From this and the harmonic map equations \eqref{harmonicmap}, the norm of the tension is found to be
\begin{equation} \label{tension}
  |\tau|^2=\frac14 \left[ \mathrm{Tr}(\operatorname{div}H + G)\right]^2 + \frac14 \mathrm{Tr} \left[(\operatorname{div}H + G)(\operatorname{div}H + G) \right]
  + \frac12 f (\operatorname{div}K)^t F (\operatorname{div}K),
\end{equation}
where
\begin{equation}
H=F^{-1}\nabla F,\quad\quad G=f^{-1}F^{-1}\left(\nabla\omega\right)^2,\quad\quad
K=f^{-1}F^{-1}\nabla\omega.
\end{equation}

\subsection{Model Map for ALE Solutions}\label{ALE sub}

\begin{figure}
\includegraphics[width=10cm]{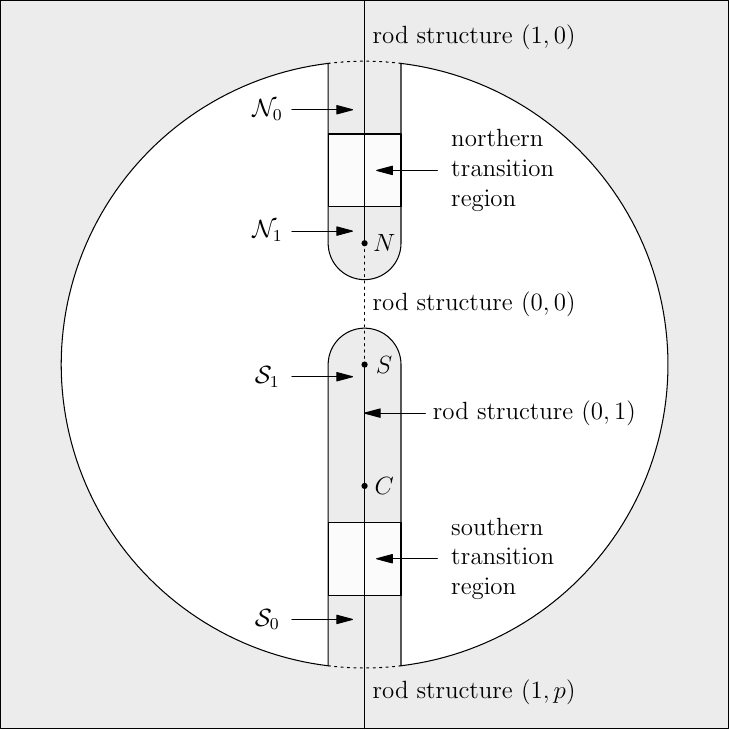}
\caption{Model Map Construction}  \label{domain}
\end{figure}

For the sake of having a specific example in mind, consider a configuration with a spherical $S^3$ horizon cross-section topology and a lens $L(p,1)$ spatial cross-section topology at infinity. The rod structure from north to south should then be $(1,0), (0,0), (0,1), (1,p)$, as in Figure \ref{domain}. The construction of the model map
differs from that in \cite{KhuriWeinsteinYamada} only in the exterior region. Thus, on the region interior to the large ball in Figure \ref{domain} all desired properties of the model map are known to be valid. Outside the large ball in Figure \ref{domain} define $\Phi_0=(F,\omega(\theta))$, where $F=F_{\text{lens}}$ is as in Section \ref{MP} and $\omega(\theta)$ is independent of $r$.
Note that since $\div (F^{-1}\nabla F)=0$, whenever $\omega$ is constant this map is harmonic because $G=K=\mathbf{0}$. Therefore $\omega$ is chosen to be the required potential constants on the intervals $[0,\epsilon]\cup [\pi-\epsilon,\pi]$, and
to smoothly connect these two constants on $[\epsilon,\pi-\epsilon]$. It follows that $|\tau(\Phi_0)|=0$ in a neighborhood of the north and south axes, and is a smooth function elsewhere in the exterior region.

It remains to show that the tension has the required fall-off at infinity.
Observe that direct computation yields
\begin{equation}
	f (\operatorname{div}K)^t F (\operatorname{div}K) =O(r^{-7}),
\end{equation}
and
\begin{equation}
	G=O(r^{-5}).
\end{equation}
The exact expressions for these quantities are similar to those in \eqref{asdf} and \eqref{;lkj} below. Although it may appear that difficulties arise due to the negative powers of $\cos(\theta/2)$ and $\sin(\theta)$, such expressions are always multiplied by derivatives of $\omega$, and since these derivatives vanish near $\theta=0,\pi$ the stated estimates hold. Now, using \eqref{tension} and the fact that $\div H=0$,
it follows that the tension decays like $O(r^{-7/2})$.

\subsection{Model Map for AKK Solutions}

Once again for the sake of having a specific example in mind, consider a rod structure $(1,0), (0,0), (0,1), (1,0)$. This corresponds to an $S^3$ horizon cross-section with  ring $S^1\times S^2$ spatial cross-sections at infinity. As above we need only give the
construction outside the large ball, as the prescription for the model map in the remaining portion of the domain is given in \cite{KhuriWeinsteinYamada}.
In the exterior region we define $\Phi_0=(F,\omega(\theta))$, where $F=F_{\text{ring}}$ is as in Section \ref{ring} and $\omega(\theta)$ is as in Section \ref{ALE sub}. Then $\div H=0$ in this region, and $\omega'=\mathbf{0}$ near the axes. Furthermore calculations show that
\begin{align}\label{asdf}
\begin{split}
	f (\operatorname{div}K)^t F (\operatorname{div}K) =&
    \frac{16\csc^4\theta \sin^6(\theta/2)}{r^9}
    \left[r \csc^6(\theta/2) \bigl(\omega_1''-(\csc\theta+2\cot\theta) \omega_1'\bigr)^2 \right. \\
    &\left. + 16 \csc^4\theta \bigl(\omega_2''+(\csc\theta-2\cot\theta) \omega_2'\bigr)^2\right]\\
    =& O(r^{-8}),
\end{split}
\end{align}
and
\begin{equation}\label{;lkj}
	G=
	\begin{pmatrix}
		 \dfrac{\omega _1'^2\cot^2(\theta/2)}{r^4} &
		 \dfrac{\omega _1'\omega _2\cot^2(\theta/2)}{r^4} \\[1ex]
		 \dfrac{\omega _1'\omega _2 \cos^2(\theta/2)}{r^3} &
		 \dfrac{\omega _2'^2 \cos^2(\theta/2)}{r^3} &
	\end{pmatrix}
	=O(r^{-3}).
\end{equation}
From \eqref{tension} it follows that $|\tau|=O(r^{-3})$.

\section{Existence and Uniqueness}
\label{existence}

With the model map in hand, the proof of Theorem \ref{thm1} may now be carried out following the now standard techniques originally developed in \cite{Weinstein}.
For the sake of completeness we sketch the arguments here. Given a rod data set $\mathcal{D}$ satisfying the hypotheses of Theorem \ref{thm1}, let $\Phi_0$ be a corresponding model map constructed in the previous section. It will be shown that there is a unique harmonic map $\Phi:\mathbb{R}^3 \setminus\Gamma\rightarrow SL(3,\mathbb{R})/SO(3)$ which is asymptotic to $\Phi_0$. Recall that two such maps are said to be asymptotic if $d(\Phi,\Phi_0)$ remains bounded near $\Gamma$, and $d(\Phi,\Phi_0)\rightarrow 0$ as $r\rightarrow \infty$, where $d(\Phi,\Phi_0)$ represents the distance in $SL(3,\mathbb{R})/SO(3)$. As is shown in \cite{KhuriWeinsteinYamada}, two maps which are asymptotic give rise to the same rod structure, and thus the spacetime resulting from the harmonic map $\Phi$ will have prescribed topology for the horizon cross-sections and spatial cross-sections at infinity, as well as prescribed angular momenta for each horizon component.

Consider first the uniqueness portion of the result. Suppose that there are two
harmonic maps $\Phi_1,\Phi_2\colon\mathbb{R}^3\setminus\Gamma \to SL(3,\R)/SO(3)$. Due to the fact that the target space is nonpositively curved, a computation yields
\begin{equation}\label{102938}
	\Delta\left( \sqrt{1 + d(\Phi_1,\Phi_2)^2} \right) \geq
	-\left( |\tau(\Phi_1)| + |\tau(\Phi_2)| \right)=0.
\end{equation}
If the two maps $\Phi_1$ and $\Phi_2$ are asymptotic to the model map $\Phi_0$, then these maps are asymptotic to each other. Therefore there is a uniform bound for the distance
$d(\Phi_1,\Phi_2)\leq C$. Since the set $\Gamma$ is of codimension 2, $\sqrt{1 + d(\Phi_1,\Phi_2)^2}$ is weakly subharmonic and the maximum principle applies \cite{Weinstein}*{Lemma 8}. As $\sqrt{1 + d(\Phi_1,\Phi_2)^2}\to 1$ at infinity, it follows that $\sqrt{1 + d(\Phi_1,\Phi_2)^2}\leq 1$. Consequently $\Phi_1=\Phi_2$.

The proof of existence proceeds as follows. Let $\Omega_\epsilon=\{x\in\R^3\colon |x|<1/\epsilon, \dist(x,\Gamma)>\epsilon\}$, and let $\Phi_\epsilon\colon\Omega_\epsilon\to SL(3,\R)/SO(3)$ be the unique harmonic map with $\Phi_\epsilon=\Phi_0$ on $\partial\Omega_\epsilon$. Due to the boundedness and decay of
$|\tau(\Phi_0)|$, there exists a positive smooth function $w$ on $\mathbb{R}^3$ satisfying $\Delta w\leq -|\tau(\Phi_0)|$ and $w\rightarrow 0$ at infinity. With the help of \eqref{102938} we then have
\begin{equation}
\Delta\left( \sqrt{1 + d(\Phi_{\epsilon},\Phi_0)^2} -w\right) \geq 0,\quad\quad\quad
\sqrt{1 + d(\Phi_{\epsilon},\Phi_0)^2} -w\leq 1\text{ }\text{ on }\text{ }\partial\Omega_{\epsilon}.
\end{equation}
The maximum principle again applies to yield
a uniform $L^\infty$ estimate for $d(\Phi_\epsilon,\Phi_0)$. This leads to a local pointwise energy estimate as in \cite{KhuriWeinsteinYamada}*{Section 6}. These bounds form the basis from which a bootstrap procedure can be employed to control all higher order derivatives of $\Phi_{\epsilon}$ on compact subsets. Hence, this sequence of maps subconverges to a harmonic map $\Phi$ asymptotic to $\Phi_0$.

\section{Conical Singularities}
\label{conical}

In this section we will prove Theorem \ref{thm2}. That is, it will be shown that there are no conical singularities on the two semi-infinite rods $\Gamma_1$ and $\Gamma_{L+1}$ for any of the solutions produced in the previous section. In particular, solutions having a single black hole and no corner points in the rod structure are void of conical singularities. This gives geometrically regular examples of black rings with AKK asymptotics, as well as black lenses with ALE asymptotics; the latter have already been exhibited in Section \ref{examples} as quotients of Myers-Perry black holes.

The absence of a conical singularity on a rod $\Gamma_l$ requires
\begin{equation}
\lim_{\rho\rightarrow 0}\frac{\rho^2 f^{-1} e^{2\sigma}}{f_{ij}u^i u^j}=1,
\end{equation}
where $u=(m_l,n_l)^t$ is the rod structure for $\Gamma_l$. This is equivalent to
\begin{equation}
b_l=2P(z)-Q(z)=0,
\end{equation}
where $b_l$ is the angle deficit which is known to be constant on each rod \cite{ChenTeo,Harmark}, and
\begin{equation}
P(z):=\lim_{\rho\rightarrow 0}\left(\sigma-\frac{1}{2}\log f\right),\quad\quad Q(z):=\lim_{\rho\rightarrow 0}\log\left(\frac{f_{ij}u^i u^j}{\rho^2}\right).
\end{equation}
Consider now the two semi-infinite rods $\Gamma_{1}$ and $\Gamma_{L+1}$. By regularity of the harmonic map and
$d(\Phi,\Phi_0)\leq C$ it may be shown that the limit $Q(z)$ exists. Moreover, since $d(\Phi,\Phi_0)\rightarrow 0$ as $r\rightarrow \infty$ we find that $Q(\infty)-Q(-\infty)=0$ as this property holds for the model map; details for this argument can be found in the proof of Theorem 11 in \cite{KhuriWeinsteinYamada}.
Observe that since
\begin{equation}
2(P(z)-P(-z))=b_1 -b_{L+1} +Q(z)-Q(-z),
\end{equation}
we have
\begin{equation}
\frac{4}{r}\int_{r/2}^r(P(z)-P(-z))dz
=b_1 - b_{L+1} +\frac{2}{r}\int_{r/2}^r (Q(z)-Q(-z))dz.
\end{equation}
Thus if it can show that
\begin{equation}\label{Plimit}
\int_{r/2}^r(P(z)-P(-z))dz=o(r)\quad\text{ as }\quad r\rightarrow \infty,
\end{equation}
then $b_1=b_{L+1}$. Furthermore, since $\sigma$ is obtained by quadrature it is only defined up to a constant, and by choosing this constant appropriately we may assume without loss of generality that $b_1=0$. The desired conclusion $b_1=b_{L+1}=0$ now follows. The rest of this section is dedicated to verifying claim \eqref{Plimit}.

Let $\gamma(\theta)=(r\sin\theta,r\cos\theta)$ be a large semi-circle connecting $\Gamma_1$ to $\Gamma_{L+1}$ in the $\rho z$-plane, so that
\begin{equation}
P(-z)-P(z)=\int_{\gamma}\left(\sigma_{\rho}-\frac{f_{\rho}}{2f}\right)d\rho
+\left(\sigma_{z}-\frac{f_{z}}{2f}\right)dz.
\end{equation}
According to \cite{IdaIshibashiShiromizu}
\begin{align}
\begin{split}
\alpha_{\rho}:=\sigma_{\rho}-\frac{f_{\rho}}{2f}=&\frac{\rho}{8}\left[
\frac{f_{\rho}^2-f_{z}^2}{f^2}+f^{ij}f^{kl}(f_{ik,\rho}f_{jl,\rho}
-f_{ik,z}f_{jl,z})+\frac{2f^{ij}}{f}(\omega_{i,\rho}\omega_{j,\rho}-\omega_{i,z}\omega_{j,z})
-\frac{4f_{\rho}}{\rho f}\right],\\
\alpha_{z}:=\sigma_{z}-\frac{f_{z}}{2f}=&\frac{\rho}{4}\left[\frac{f_{\rho}f_{z}}{f^2}
+f^{ij}f^{kl}f_{ik,\rho}f_{jl,z}+\frac{2f^{ij}}{f}\omega_{i,\rho}\omega_{j,z}
-\frac{2f_{z}}{\rho f}\right],
\end{split}
\end{align}
and therefore
\begin{align}\label{l}
\begin{split}
\int_{r/2}^r (P(-z)-P(z))dz=&\int_{r/2}^{r}\int_{0}^{\pi}
(\alpha_{\rho}\cos\theta -\alpha_{z}\sin\theta)rd\theta dr\\
=&\frac{1}{2\pi}\int_{B_r\setminus B_{r/2}}\left(\alpha_{\rho}
\frac{\cos\theta}{\rho}-\alpha_{z}\frac{\sin\theta}{\rho}\right)dx.
\end{split}
\end{align}
It turns out that this integral may be estimated in terms of a reduced (or renormalized)
energy, which will be shown to have the appropriate asymptotics. Consider first the $\rho$-term, and observe that it may be re-expressed as
\begin{align}\label{alpha}
\begin{split}
8\frac{\alpha_{\rho}}{\rho}=&4\overline{\mathcal{E}}(\Phi)
-2(\partial_{z}\log f)^2
-(f^{ij}f_{ik,z}-f^{0ij}f^{0}_{ik,z})(f^{kl}f_{lj,z}-f^{0kl}f^{0}_{lj,z})
-4\frac{f^{ij}}{f}\omega_{i,z}\omega_{j,z}\\
&-f^{ij}f^{kl}f_{ik,z}f_{jl,z}
+2f^{0ij}f^{0}_{ik,\rho}(f^{kl}f_{lj,\rho}-f^{0kl}f^{0}_{lj,\rho})
+\left(f^{0ij}f^{0kl}f_{ik,\rho}^{0}f_{jl,\rho}^{0}
-|\nabla\log\rho^2|^2\right),
\end{split}
\end{align}
where the energy density and reduced energy density are given by
\begin{equation}
\mathcal{E}(\Phi)=\frac{1}{4}|\nabla\log f|^2
+\frac{1}{4}f^{ij}f^{kl}\nabla f_{ik}\cdot\nabla f_{jl}
+\frac{1}{2}\frac{f^{ij}}{f}\nabla\omega_{i}\cdot\nabla\omega_{j},
\end{equation}
\begin{equation}
\overline{\mathcal{E}}(\Phi)=\frac{1}{4}|\nabla(\log f-\log\rho^2)|^2
+\frac{1}{4}(f^{ij}\nabla f_{ik}-f^{0ij}\nabla f^{0}_{ik})\cdot
(f^{kl}\nabla f_{lj}-f^{0kl}\nabla f^{0}_{lj})
%&+\frac{1}{4}(f^{ij}f_{ik,z}-f^{0ij}f^{0}_{ik,z})
%(f^{kl}f_{lj,z}-f^{0kl}f^{0}_{lj,z})
+\frac{1}{2}\frac{f^{ij}}{f}\nabla\omega_{i}\cdot\nabla\omega_{j}.
\end{equation}

All the terms on the first line of \eqref{alpha} are part of the reduced energy density. Let us now estimate the last term on the second line. In the AKK case the model map matrix $F^0=(f^{0}_{ij})$ is diagonal \eqref{ringF}, and therefore a computation yields
\begin{align}
\begin{split}
f^{0ij}f^{0kl}f_{ik,\rho}^{0}f_{jl,\rho}^{0}
=&(\partial_{\rho}\log f_{11}^0)^2+(\partial_{\rho}\log f_{22}^0)^2\\
=&\left(\frac{2}{\rho}+\frac{\cos\theta_{a}-\cos\theta_{-a}}{\rho}\right)^2
=\frac{4}{\rho^2}-\frac{4a\sin^{2}\theta}{r\rho^2}+O\left(\frac{\sin\theta}{r^2\rho^2}\right).
\end{split}
\end{align}
In the ALE case, although $F^0$ is not necessarily diagonal \eqref{ALEF}, $\tilde{F}^0=h^{-1}F^0 (h^t)^{-1}$ is diagonal so that
\begin{align}
\begin{split}
f^{0ij}f^{0kl}f_{ik,\rho}^{0}f_{jl,\rho}^{0}=&
\tilde{f}^{0ij}\tilde{f}^{0kl}\tilde{f}_{ik,\rho}^{0}\tilde{f}_{jl,\rho}^{0}\\
=&(\partial_{\rho}\log \tilde{f}_{11}^0)^2+(\partial_{\rho}\log \tilde{f}_{22}^0)^2\\
=&\left(\frac{1+\cos\theta}{\rho}\right)^2
+\left(\frac{1-\cos\theta}{\rho}\right)^2
=\frac{4}{\rho^2}-\frac{2\sin^{2}\theta}{\rho^2}.
\end{split}
\end{align}
It follows that in both cases
\begin{equation}\label{rhoremainder}
\int_{B_r\setminus B_{r/2}}\left(f^{0ij}f^{0kl}f_{ik,\rho}^{0}f_{jl,\rho}^{0}
-|\nabla\log\rho^2|^2\right)\cos\theta dx
=O\left(\frac{1}{r}\right),
\end{equation}
since the terms involving $\sin^{2}\theta$ integrate to zero against $\cos\theta$.

The first term on the second line of \eqref{alpha} may be rewritten as
\begin{align}
\begin{split}
f^{ij}f^{kl}f_{ik,z}f_{jl,z}=&
(f^{ij}f_{ik,z}-f^{0ij}f^{0}_{ik,z})
(f^{kl}f_{lj,z}-f^{0kl}f^{0}_{lj,z})\\
&+2f^{0ij}f^{0}_{ik,z}(f^{kl}f_{lj,z}-f^{0kl}f^{0}_{lj,z})
+f^{0ij}f^{0kl}f_{ik,z}^{0}f_{jl,z}^{0}.
\end{split}
\end{align}
Moreover, in a similar manner to the above calculations we find that in the AKK case
\begin{equation}\label{alsk0}
f^{0ij}f^{0kl}f_{ik,z}^{0}f_{jl,z}^{0}
=(\partial_{z}\log f_{11}^0)^2+(\partial_{z}\log f_{22}^0)^2
=\left(\frac{1}{r_a}-\frac{1}{r_{-a}}\right)^2
=\frac{a^2\cos^{2}\theta}{r^4}+O\left(\frac{1}{r^5}\right),
\end{equation}
while in the ALE setting
\begin{equation}\label{alsk1}
f^{0ij}f^{0kl}f_{ik,z}^{0}f_{jl,z}^{0}=
\tilde{f}^{0ij}\tilde{f}^{0kl}\tilde{f}_{ik,z}^{0}\tilde{f}_{jl,z}^{0}
=(\partial_{z}\log \tilde{f}_{11}^0)^2+(\partial_{z}\log \tilde{f}_{22}^0)^2
=\frac{2}{r^2}.
\end{equation}
Hence in both cases
\begin{equation}\label{zremainder}
\int_{B_r\setminus B_{r/2}}f^{0ij}f^{0kl}f_{ik,z}^{0}f_{jl,z}^{0}
\cos\theta dx
=O\left(\frac{1}{r^2}\right),
\end{equation}
since the term on the right-hand side of \eqref{alsk0} and \eqref{alsk1} integrates to zero against $\cos\theta$.
Combining the above computations with \eqref{alpha} yields
\begin{align}\label{u}
\begin{split}
\left|\int_{B_{r}\setminus B_{r/2}}\alpha_{\rho}\frac{\cos\theta}{\rho}dx\right|
\leq& 2\int_{B_{r}\setminus B_{r/2}}\overline{\mathcal{E}}(\Phi)dx
+\frac{1}{4}\left|\int_{B_{r}\setminus B_{r/2}}f^{0ij}f^{0}_{ik,\rho}(f^{kl}f_{lj,\rho}-f^{0kl}f^{0}_{lj,\rho})\cos\theta dx\right|\\
&
+\frac{1}{4}\left|\int_{B_{r}\setminus B_{r/2}}f^{0ij}f^{0}_{ik,z}(f^{kl}f_{lj,z}-f^{0kl}f^{0}_{lj,z})\cos\theta dx\right|
+O\left(\frac{1}{r}\right).
\end{split}
\end{align}

The reduced energy may be estimated in two possible ways. One method exploits the fact that the target symmetric space $SL(3,\mathbb{R})/SO(3)$ for the harmonic map has nonpositive curvature, and therefore the energy is naturally convex along geodesic deformations. Then by connecting the harmonic map $\Phi$ to its model map $\Phi_0$ via a geodesic in the target space, and using that the energy is a convex function of the geodesic parameter, it can be shown that the reduced energy of $\Phi$ is dominated by the reduced energy of $\Phi_0$. This provides the desired bounds for the reduced energy of $\Phi$. However, some care must be taken to implement this procedure since the energy of $\Phi$ (and of $\Phi_0$) is infinite. In order to prove that the reduced energy inherits the convexity property from the pure energy, a cut-off argument must be used near the axes. Such a procedure has been carried out successfully within the context of mass-angular momentum inequalities, see for example \cite{KhuriWeinstein,AlaeeKhuriKunduri,AlaeeKhuriKunduri1,AlaeeKhuriKunduri2}.

An alternate approach to estimating the reduced energy, which is more straightforward to carry out, consists of applying standard `energy' methods for obtaining local a priori estimates associated with elliptic PDE. This entails multiplying the Euler-Lagrange equations by appropriate functions and integrating by parts. For example, consider the following equation arising from the harmonic map equations \eqref{harmonicmap} and the fact that $\log\rho$ is a harmonic function
\begin{equation}\label{z}
\Delta(\log f-\log\rho^2)=-\frac{f^{ij}}{f}\nabla\omega_{i}\cdot\nabla\omega_{j}.
\end{equation}
Let $\chi$ be a smooth cut-off function with $\mathrm{supp}\!\text{ }\chi\subset B_{2r}\setminus B_{r/4}$, multiply \eqref{z} by $\chi^2(\log f-\log\rho^2)$, and integrate to find
\begin{align}
\begin{split}
\int_{B_{2r}\setminus B_{r/4}}\chi^2|\nabla(\log f-\log\rho^2)|^2 dx
=&\int_{B_{2r}\setminus B_{r/4}}\chi^2 (\log f-\log\rho^2)\frac{f^{ij}}{f}\nabla\omega_{i}\cdot\nabla\omega_{j}dx\\
&
-2\int_{B_{2r}\setminus B_{r/4}}\chi(\log f-\log\rho^2)\nabla\chi\cdot\nabla(\log f-\log\rho^2)dx.
\end{split}
\end{align}
It follows that
\begin{align}
\begin{split}
\int_{B_{2r}\setminus B_{r/4}}\!\!\!\!\!\!\!\!\!\chi^2|\nabla(\log f-\log\rho^2)|^2 dx
\leq & \sup_{B_{2r}\setminus B_{r/4}}|\log f-\log\rho^2|\int_{B_{2r}\setminus B_{r/4}}\!\!\!\!\!\!\!\!\!|\nabla\chi|^2 dx\\
&+2\sup_{B_{2r}\setminus B_{r/4}}|\log f-\log\rho^2|\int_{B_{2r}\setminus B_{r/4}}\!\!\!\!\!\!\!\!\!\chi^2 \frac{f^{ij}}{f}\nabla\omega_{i}\cdot\nabla\omega_{j}dx.
\end{split}
\end{align}
Similarly, multiplying the harmonic map equation for $\omega_i$ by $\chi^2 f^{-1}f^{ij}(\omega_{j}-\omega_{j}^0)$ and integrating produces
\begin{equation}
\int_{B_{2r}\setminus B_{r/4}}\chi^2 \frac{f^{ij}}{f}\nabla\omega_{i}\cdot\nabla\omega_{j}dx
\leq c\sup_{B_{2r}\setminus B_{r/4}}\left[\frac{f^{ij}(\omega_{i}-\omega_{i}^0)(\omega_{j}-\omega_{j}^0)}{f}\right]
\int_{B_{2r}\setminus B_{r/4}}|\nabla\chi|^2 dx.
\end{equation}
Therefore
\begin{align}\label{k}
\begin{split}
&\int_{B_{2r}\setminus B_{r/4}}\!\!\!\!\!\!\!\!\chi^2|\nabla(\log f-\log\rho^2)|^2 dx\\
\leq & c\sup_{B_{2r}\setminus B_{r/4}}|\log f-\log\rho^2|\left[1
+\sup_{B_{2r}\setminus B_{r/4}}\left(\frac{f^{ij}(\omega_{i}-\omega_{i}^0)(\omega_{j}-\omega_{j}^0)}{f}\right)\right]
\int_{B_{2r}\setminus B_{r/4}}\!\!\!\!\!\!\!\!|\nabla\chi|^2 dx.
\end{split}
\end{align}
Since $d(\Phi,\Phi_0)\rightarrow 0$ as $r\rightarrow \infty$ the term in brackets on the right-hand side of \eqref{k} decays at infinity. Thus by choosing a cut-off function $\chi\equiv 1$ on $B_{r}\setminus B_{r/2}$ which vanishes outside $B_{2r}\setminus B_{r/4}$, so that $|\nabla\chi|\sim 1/r$, we have
\begin{equation}
\int_{B_{r}\setminus B_{r/2}}|\nabla(\log f-\log\rho^2)|^2 dx=o(r).
\end{equation}

Similar considerations may be used to estimate all remaining terms in the reduced energy, as well as the two other terms on the right-hand side of \eqref{u}. It follows that
\begin{equation}
\left|\int_{B_{r}\setminus B_{r/2}}\alpha_{\rho}\frac{\cos\theta}{\rho}dx\right|
=o(r).
\end{equation}
Moreover, an analogous procedure yields
\begin{equation}
\left|\int_{B_{r}\setminus B_{r/2}}\alpha_{z}\frac{\sin\theta}{\rho}dx\right|
=o(r).
\end{equation}
This together with \eqref{l} gives the desired conclusion \eqref{Plimit}.

%[[Reduction argument needs Chrusciel Galloway result for potentials. Actually %Hollands orbit space theorem still applies to show that the orbit space is simply %connected.]]

%The lens can be considered as a KK asymptotic with twisted circle bundle fibration over %$S^2$, by Hopf fibration.

%Also, the lens can be considered a generalization of KK since typically one can replace %the role of a torus by any compact manifold. So Minkowski times a lens is asymptotically %KK.

\end{document}